\documentclass[submission, Phys]{SciPost}

\usepackage{ulem}
\usepackage[english]{babel}
\usepackage{amsmath}
\usepackage{graphicx}
\usepackage{amsfonts}
\usepackage{amssymb}
\usepackage{graphicx}
\usepackage{mathrsfs}
\usepackage{layout}
\usepackage{hyperref}
\usepackage{mathtools}
\usepackage{url}
\usepackage{titlesec}
\usepackage{bbold}
\usepackage{caption}
\usepackage{subcaption}
\usepackage{lineno}

\usepackage{array}

\newcommand{\dd}{\text{d}}

\newcommand{\p}{\partial}
\newcommand{\vx}{\mathbf{x}}
\newcommand{\vxp}{\vx'}
\newcommand{\vq}{\mathbf{q}}

\newcommand{\vp}{\mathbf{p}}
\newcommand{\vv}{\mathbf{v}}

\newcommand{\bla}{\big\langle}
\newcommand{\bra}{\big\rangle}
\newcommand{\la}{\langle}
\newcommand{\ra}{\rangle}

\newcommand\omext{\varpi}
\newcommand\omint{\omega}
\newcommand\ptwo{\widetilde{p}}

\newcommand\omtwo{\widetilde\omext}

\newcommand\hp{\hat{p}}
\newcommand\homext{\hat{\omext}}
\newcommand\hpexit{\hp_\text{exit}}
\newcommand\homint{\hat{\omint}}
\newcommand\hq{\hat{q}}

\newcommand\hQ{\hat{Q}}
\newcommand\hOmint{\hat{\Omega}}

\begin{document}

\begin{center}
{\Large \textbf{
Unveiling the different scaling regimes }}

\vspace{0.2cm}
{\Large \textbf{of the one-dimensional Kardar-Parisi-Zhang--Burgers equation}}

\vspace{0.2cm}
{\Large \textbf{using the functional renormalisation group
}}
\end{center}

\begin{center}
Liubov Gosteva\textsuperscript{1},
Nicol\'as Wschebor \textsuperscript{2},
L\'eonie Canet\textsuperscript{1*}
\end{center}

\begin{center}
{$^1$} University Grenoble Alpes, CNRS, LPMMC, 38000 Grenoble, France\\
{$^2$} Instituto de F\'isica, Facultad de Ingenier\'ia, Universidad de la Rep\'ublica, J.H.y Reissig 565, 11000 Montevideo, Uruguay\\
{$^*$} {\small leonie.canet@lpmmc.cnrs.fr}
\end{center}

\begin{center}
\begin{minipage}{14cm}
{\small
The Kardar-Parisi-Zhang (KPZ) equation is a celebrated non-linear stochastic equation featuring non-equilibrium scaling. Although in one dimension, its statistical properties are very well understood, a new scaling regime has been reported in recent numerical simulations. This new regime is characterised by a dynamical exponent $z=1$, markedly different from the expected one $z=3/2$ for the KPZ universality class, and it emerges when approaching the inviscid limit. The origin of this scaling has been traced down to the existence of a new fixed point, termed the inviscid Burgers (IB) fixed point, which was uncovered using the functional renormalisation group (FRG).
The FRG equations  can be solved analytically in the asymptotic regime of vanishing viscosity and large momenta, showing that indeed $z=1$ exactly at the IB fixed point. In this work, we set up an advanced method  to numerically solve the full FRG flow equations in a certain approximation, which allows us to determine in a unified way the correlation function over the whole range of momenta, not restricted to some particular regime. We analyse the crossover between the different fixed points, and quantitatively determine the extent of the IB regime.
}
\end{minipage}
\end{center}

\vspace{5pt}

\section{Introduction}
\label{sec:intro}

Simplified hydrodynamical equations play a pivotal role to address fundamental problems of fluid mechanics such as turbulence. The one-dimensional (1D) Burgers equation~\cite{Burgers48} stands as one of them and has been widely studied both from the physical and the mathematical perspectives~\cite{Bec2007}. When a thermal noise is added to the Burgers equation, as in the seminal papers by Forster, Nelson, Stephen~\cite{Forster76,Forster77}, it maps  onto the now celebrated  Kardar-Parisi-Zhang (KPZ) equation~\cite{Kardar86}, which was derived in a completely different context, namely to describe the stochastic growth and roughening of an interface. The dynamics of the height field $h(t,\vx)$ is given by
\begin{equation}
\label{eq:KPZ}
    \partial_t h(t,\vx) = \nu \nabla^2 h(t,\vx) + \frac{\lambda}{2}(\nabla h(t,\vx))^2 + \sqrt{D} \eta(t,\vx)\,,
\end{equation}
from which one recovers the stochastic Burgers equation setting $\vv=-\lambda \nabla h$.
In this equation, $\nu$ is the surface tension (or viscosity for the fluid), $\lambda$ is the strength of the non-linearity, and  $\eta(t,\vx)$ is a Gaussian noise of amplitude $\sqrt{D}$ with zero mean and covariance
\begin{equation}
\label{eq:etaKPZ}
\bla \eta(t,\vx) \eta(t',\vxp)\bra = 2 \delta(t-t')\delta^d(\vx-\vxp)\,,
\end{equation}
where $d$ is the space dimension. Time and field can be rescaled such that there remains a unique non-linear coupling parameter $g=\lambda^2 D/\nu^3$ (or alternatively the Reynolds number for the fluid).
 A wealth of exact results is available on the statistical properties of the one-dimensional (1D) KPZ equation~\cite{Corwin12,Takeuchi18}.
 In particular, it is characterised by a super-diffusive scaling: the correlation time $\tau$ decays with the wavenumber $k$ as $\tau(k)\sim k^{-z}$ with $z=3/2$, where $z$ is called the dynamical critical exponent.

 On the other hand, in the inviscid limit, the Burgers equation is known to exhibit a completely different behaviour, featuring shocks, or {\it tygers}, which are transient localised oscillatory structures appearing in the spectrally (Galerkin)-truncated version of his equation, yielding in both cases a very distinct scaling behaviour~\cite{Majda2000}. Since an exact mapping  connects the two equations, this raises the interesting question of how these two realms are related together. This was first investigated numerically in  Refs.~\cite{Brachet2022,Rodriguez2022}, which showed that, not only in the inviscid limit, but also at small but finite viscosity, a new scaling with $z=1$ is present in the KPZ-Burgers equation.  This scaling, which emerges at large momenta (UV), has eluded from the exact results of the KPZ equation. The origin of this scaling has been resolved using  functional renormalisation group (FRG), which revealed the existence of a new fixed point of the KPZ equation, called inviscid Burgers (IB) fixed point. This fixed point corresponds to zero viscosity (or equivalently infinite  non-linearity $g$), and it features a dynamical exponent $z=1$~\cite{Fontaine2023InvBurgers, Gosteva2024}. It is unstable, {\it i.e.} it is a UV fixed point, and thus it influences  large momenta, while the infrared regime (IR)  is controlled by the  KPZ fixed point associated with $z=3/2$.

 The existence of the IB fixed point was demonstrated in~\cite{Fontaine2023InvBurgers, Gosteva2024} within a simple approximation of the FRG, which is suitable to describe  the limit of small wavenumbers compared to the renormalisation group scale (but which can be large compared to the IR KPZ scale). It will be referred to as `small-$p$' approximation in the following. Besides, another FRG approach was developed, referred to as `large-$p$' approximation in the following, to access the limit of large wavenumbers. This approach allows one to derive the exact result $z=1$ at the IB fixed point. However, it does not enable one to describe the IR regime at the same time, such that the KPZ fixed point is not visible in the large-$p$ approximation.
  Thus, the crossover between the KPZ and IB fixed points has not been quantitatively investigated yet. The purpose of this work is to bridge this gap.
  For this, we develop an advanced numerical method to couple and solve simultaneously the small-$p$ and the large-$p$ FRG equations. This permits us to access in a unified way the different scaling regimes of the 1D KPZ-Burgers equation, and to quantify the extent of the IB regime.

The remainder of this paper is organised as follows. In Sec.~\ref{sec:FRG}, we briefly review the FRG formalism, and present the small-$p$ and large-$p$ approximations. While the resulting flow equations were already obtained in previous works, we devise in Sec.~\ref{sec:num} a numerical scheme to integrate simultaneously the two sets of flow equations, which is the main contribution of this paper. We present in Sec.~\ref{sec:res} the results obtained using this scheme.

\section{Functional renormalisation group}
\label{sec:FRG}

\subsection{FRG formalism}

To introduce the FRG formalism, let us firstly rewrite the KPZ equation in a field-theoretical form~\cite{Martin73,Janssen76,Dominicis76}
\begin{align}
    \mathcal{Z}[j,\bar{j}] &= \int \mathcal{D} h \mathcal{D} \bar{h}\, e^{-{\cal S}[h,\bar{h}] + \int_{t,\vx}(j h + \bar j \bar{h})}\, ,\nonumber\\
	\mathcal{S}[h,\bar{h}] &= \int_{t,\vx}\Big\{\bar{h} \Big[\p_t h -\frac \lambda 2 (\nabla h)^2 - \nu \nabla^2 h   \Big] - D\,\bar{h}^2\Big\}\,,
\label{eq:action}
\end{align}
where $\bar{h}(t,\vx)$ is the response field, and $j(t,\vx),\bar j(t,\vx)$ are the sources.
  The shorthand notations $\int_{t} \equiv \int_{-\infty}^{+\infty} \dd t$, $\int_{\vx} \equiv \int_{\mathbb{R}^d} \dd^d\vx$, $\int_\omint \equiv \int_{-\infty}^{+\infty} d\omint/(2\pi)$, $\int_\vq \equiv \int_{\mathbb{R}^d} d^d\vq/(2\pi)^d$ are used throughout this paper.
The FRG is conceived as a powerful implementation of Wilson's RG procedure, driven by the idea  of progressively averaging over the fluctuation modes~\cite{Berges2002,Dupuis2021}. To achieve this, one adds to the action in the generating functional ${\cal Z}$ a quadratic term $\Delta \mathcal{S}_\kappa[h,\bar{h}] = \int_{t,\vx,\vx'}\left( - \bar{h}(t,\vx) R^\nu_\kappa(\vx-\vx') \nabla'^2 h(t,\vx') - \bar{h}(t,\vx) R^D_\kappa(\vx-\vx') \bar{h}(t,\vx')  \right)$ where $\kappa$ is the RG scale, and $R^{X}_\kappa(x)$, with $X=\nu,D$, are the regulators satisfying the following general properties. It is large $R^X_\kappa(q)\sim \kappa^2$ at small momenta $q\lesssim\kappa$ and it vanishes $R^X_\kappa(q)\simeq 0$ at large $q$. Thus, the introduced quadratic terms play the role of a large  ``mass'' for slow modes such that they do not contribute to the functional integration in ${\cal Z}$, while they leave the fast modes unaffected such that they are averaged over.

Therefore, by changing the RG scale from the microscopic scale $\kappa=\Lambda$ (large momenta, short distances) to the macroscopic one $\kappa=0$, one integrates the fluctuations in ${\cal Z}$ progressively, scale by scale. In the presence of $\Delta \mathcal{S}_\kappa$,  the generating functional becomes $\kappa$-dependent and is denoted  as $\mathcal{Z}_\kappa$.
One then defines the effective average action $\Gamma_\kappa$ as $\Gamma_\kappa = -\ln\mathcal{Z}_\kappa + \int_{t,\vx} \big(j \la h \ra + \bar{j} \la \bar{h} \ra\big) - \Delta \mathcal{S}_\kappa[\la h \ra, \la \bar{h} \ra]$. By construction, this object coincides with the microscopic action $\mathcal{S}$ at $\kappa=\Lambda$ (since  all fluctuations are then suppressed), while it contains all the statistical properties of the system at $\kappa=0$ (since all fluctuations are then included).
The evolution of $\Gamma_\kappa$ with the RG scale is described by the Wetterich equation~\cite{Wetterich93,Ellwanger94,Morris94}
\begin{equation}\label{eq:Wetterich}
    \partial_{\kappa} \Gamma_{\kappa} =
    \frac{1}{2}\textrm{tr}\,\int_{\omega,\vq}
    \partial_{\kappa} \mathcal{R}_{\kappa}\,
    \mathcal{G}_{\kappa}\,,\qquad \hbox{with}\quad \mathcal{G}_{\kappa} \equiv \left(
    \Gamma_{\kappa}^{(2)} + \mathcal{R}_{\kappa}
\right)^{-1}\,,
\end{equation}
where the  trace means summation over all fields,
$\mathcal{G}_{\kappa}$  is the propagator matrix, $\mathcal{R}_{\kappa}$  is the regulator matrix with elements $q^2R^\nu_\kappa(q)$ and $-2R^D_\kappa(q)$, and $\Gamma_{\kappa}^{(2)}$ the Hessian
of $\Gamma_\kappa$. This is an exact RG equation, but it cannot be solved without approximations in general since it yields a coupled hierarchy of flow equations for $n$-point vertex functions $\Gamma_\kappa^{(n)}$ which is not closed. One thus has to resort to some approximations.

Different approximation schemes within this formalism are available and have been thoroughly studied, we refer the reader to standard reviews for a detailed account~\cite{Berges2002,Delamotte2012,Dupuis2021}.  In this work, we use two complementary approximations: one suitable to study the small-momentum (IR) domain (small-$p$ approximation), and the other one devised to access the large-momentum (UV) domain (large-$p$ approximation). Both approximations were already used in the context of the KPZ--Burgers equation, and the resulting flow equations were derived in previous works~\cite{Canet2010,Canet2011kpz,Kloss2012,Fontaine2023InvBurgers}. We briefly present them in the next subsections. While these equations were studied so far separately, the originality of this work is to combine them together within a dedicated numerical scheme, in order to obtain a unified picture across all momentum scales, which we present in Sec.~\ref{sec:num}.

\subsection{Flow equations for the small-$p$ regime}

For the small-$p$ region, we use the next-to-leading-order (NLO) approximation introduced in~\cite{Kloss2012}. This approximation consists in considering an ansatz for $\Gamma_\kappa$, which has the same form as the KPZ action~\eqref{eq:action} but where the microscopic (bare) parameters of the KPZ equation $\nu$, $D$, $\lambda$ are promoted to effective renormalised functions $f^{\nu}_\kappa(\omext,p)$, $f^{D}_\kappa(\omext,p)$, $f^{\lambda}_\kappa(\omext,p)$, which depend on momentum $p\equiv|\vp|$ and frequency $\omext$. In fact, in one dimension, the time-reversal symmetry of the KPZ action, expressed within this approximation, imposes that $f^\lambda_\kappa(\omext,p)\equiv \lambda$, which means that the interaction is not renormalised.  One can introduce a renormalised viscosity and noise amplitude as $\nu_\kappa = f^\nu_\kappa(0,0)$ and $D_\kappa = f^D_\kappa(0,0)$, from which
 anomalous dimensions are defined as $\eta^\nu_\kappa \equiv  -\p_s \ln \nu_\kappa$ and $\eta^D_\kappa \equiv  -\p_s \ln D_\kappa$. In one dimension, the time-reversal symmetry  also enforces the relation $f^\nu_\kappa(\omext,p) / \nu = f^D_\kappa(\omext,p) / D$, such that there remains only one independent function $f^{\nu, D}_\kappa \equiv  f_\kappa$ and one anomalous dimension $\eta^{\nu,D}_\kappa\equiv \eta_\kappa$. To preserve this symmetry along the flow, one can choose the regulators as $R^X_\kappa(q) = X_\kappa \hat r(y)$ with $y\equiv q^2/\kappa^2$. In this work we use the Wetterich regulator $r(y)=2/(\exp(y)-1)$.
 One can then project the exact Wetterich equation~\eqref{eq:Wetterich} onto this ansatz to deduce the flow equations for the function $f_\kappa$ and for $\eta_\kappa$.

In order to study fixed points of the RG flow, it is  necessary to express it in terms of dimensionless variables $\hp \equiv p/\kappa$, $\homext \equiv \omext/(\nu_\kappa \kappa^2)$, $\hat f_\kappa(\homext,\hp) \equiv f_\kappa(\omext,p) / \nu_\kappa$. One introduces  the dimensionless coupling parameter $\hat g_\kappa =  \kappa^{d-2} \lambda^2 D_\kappa/\nu_\kappa^3$.
 The NLO flow equation  for the function $\hat f_\kappa$ reads in 1D~\cite{Kloss2012,Fontaine2023InvBurgers}
\begin{equation}
\p_s \hat f_\kappa(\homext,\hp) = (\eta_\kappa + \hp\p_{\hp} + (2-\eta_\kappa)\homext\p_{\homext})\hat f_\kappa(\homext,\hp) + \hat J_\kappa(\homext,\hp)\,,
\label{eq:NLOdimless}
\end{equation}
where $s=\ln (\kappa/\Lambda)$, and
\begin{equation}
\label{eq:Jnlo}
\hat J_\kappa(\homext,\hp) = 
2\hat{g}_\kappa \int_{\homint,\hq} 
\frac{(\hq^2+\hp\hq)^2 \hat{k}_\kappa(\hOmint,\hQ)}{\hat{P}_\kappa(\hOmint,\hQ)\hat{P}_\kappa(\homint,\hq)} 
\left( 
	1 - \frac{2\hq^4 \hat{k}_\kappa(\homint,\hq)^2}{\hat{P}_\kappa(\homint,\hq)}
\right) \widehat{\p_s R}_\kappa(\hq) \,,
\end{equation}
with $\widehat{\p_s R}_\kappa(\hq) \equiv -(\eta_\kappa+2y\p_y) \hat r(y)$, 
$\hQ \equiv \hp+\hq$, $\hOmint \equiv \homext+\homint$, $\hat{k}_\kappa(\homint,\hq) \equiv \hat f_\kappa(\homint,\hq) + \hat r(\hq^2)$, $\hat{P}_\kappa(\homint,\hq) \equiv \homint^2 + (\hq^2 \hat{k}_\kappa(\homint,\hq))^2$.
 Note that an additional approximation is implemented at NLO, which is to neglect the $\hat\omint$-dependence of $\hat f_\kappa$  in \eqref{eq:Jnlo}. Thus it amounts to replacing $\hat{k}_\kappa(\homint,\hq)$ and $\hat{k}_\kappa(\hOmint,\hq)$  by $\hat{k}_\kappa(0,\hq)$ in the integrand.
This equation together with the equations for $\hat g_\kappa$ and $\eta_\kappa$
\begin{equation}
\p_s \hat g_\kappa = \hat g_\kappa (2\eta_\kappa - 1), \quad \eta_\kappa + \hat J_\kappa(0,0)= 0
\label{eq:NLOeta}
\end{equation}
provide the small-$p$ RG flow, which gives in this regime the variation of the system under a change of scale, starting from the microscopic initial condition $\hat f_\Lambda(\homext,\hp) = 1$ and some given value of $\hat g_\Lambda$ at the microscopic scale $\kappa=\Lambda$. It is important to stress that here `small-$p$' means small compared to $\kappa$. The momentum $p$ can nevertheless be large compared to the physical IR scale.

The procedure to obtain the numerical solution of these equations is described in Sec.~\ref{sec:num}. One can then integrate the flows of the effective noise $D_\kappa$ and  viscosity $\nu_\kappa$ from their definitions and the calculated $\eta_\kappa$~(\ref{eq:NLOeta}).
Thus, once the equation~(\ref{eq:NLOdimless}) is solved and $\hat f_\kappa(\homext,\hp)$ is found for all $\kappa$ from $\Lambda$ to some small value corresponding to the IR regime, one can reconstruct  the dimensionful functions $f_\kappa^{\nu,D}(\omext,p)$ and compute the connected correlation function
$C(\omext, p) \equiv \mathcal{F}\left[ \big\la h(t,x)h(0,0) \big\ra_c \right]$, where $\mathcal{F}$ denotes the Fourier transform and $\la...\ra$ the statistical average, as~\cite{Kloss2012}
\begin{equation}
C_\kappa(\omext, p) = \frac{2f_\kappa^D(\omext, p)}{\omext^2 + (p^2f_\kappa^\nu(\omext, p))^2}\,.
\label{eq:Corr}
\end{equation}
It was shown in Refs.~\cite{Fontaine2023InvBurgers,Gosteva2024} that, besides the IR KPZ fixed point and the UV EW fixed point,  the flow equation~\eqref{eq:NLOeta} features another UV fixed point, the IB fixed point, which is strong-coupling as it corresponds to the limit of vanishing viscosity, {\it i.e} infinite coupling $\hat{g}_\kappa$. It is further studied in Sec.~\ref{sec:res}.

\subsection{Flow equations for the large-$p$ regime}

In the  large-$p$ limit, it  has been shown that for some problems, including the Navier-Stokes and the Burgers equations, the FRG flow equations for all $n$-point correlation functions can be closed without other assumptions than the limit of large momenta  and the existence of a scaling regime~\cite{Tarpin2018,Canet2022,Fontaine2023InvBurgers}. This remarkable closure originates in the presence of the regulator and in the existence of (extended) symmetries.  Here again, large-$p$ means that the momenta are large compared to the RG scale $\kappa$. Indeed, the properties of the regulator allows one to separate in a controlled manner small and large momentum modes in the flow equation. The symmetries of the Navier-Stokes or KPZ-Burgers actions provide a set of exact Ward identities for arbitrary vertex functions in  specific momentum configurations,  which are precisely the ones occurring in the flow equations in the limit of large momenta. From these two ingredients together stems an exact closure of the flow equations in this specific limit. For the two-point correlation function of the 1D KPZ-Burgers equation, the large-$p$ flow equation simply reads~\cite{Fontaine2023InvBurgers}
\begin{equation}
\p_s C_{\kappa}(\varpi, p) =  {p^2}
\int_{\omega} \dfrac{C_{\kappa}(\omega+\varpi,  p) -
C_{\kappa}(\varpi,  p)}{\omega^2}\,  {\cal J}_\kappa(\omega)\label{eq:flowC}\,,\qquad
{\cal J}_\kappa(\omega)\equiv  \int_{q} q^2\, \tilde{\p}_sC_{\kappa}(\omega, q)\,,
\end{equation}
where the operator $\tilde{\p}_s$ only acts on the $s$-dependence of the regulators.
Fourier transforming back in time delays, this equation can be solved at the fixed point, yielding the general solution~\cite{Fontaine2023InvBurgers}:
\begin{equation}\label{solutionCasym}
    C(t,p) =  C(0,p){\times}
    \begin{cases}
        \exp\left( - \mu_0 (pt)^2 \right),\quad t\ll \tau_c
        \\
        \exp\left( - \mu_{\infty} p^2 |t| \right),\quad t\gg \tau_c
    \end{cases}
\end{equation}
where $\tau_c$ is a characteristic time scale, and $\mu_0$, $\mu_{\infty}$ are non-universal constants related to ${\cal J}_\kappa$, whose determination requires the full integration of the flow. The first line of~\eqref{solutionCasym} shows the exact result $z=1$ at large momenta, which are controlled by the IB fixed point as shown in Sec.~\ref{sec:res}. However,  the quantity ${\cal J}_\kappa$ is dominated by small momenta $q\lesssim \kappa$ since the integral is cutoff to $q\lesssim \kappa$ due to the presence of the regulator. The large-$p$ equation is thus not appropriate to compute it. One should use instead the small-$p$ equation, and couple it to the large-$p$ one.

This is the purpose of this work.  This will give us access to the explicit form of the correlation function in all momentum regimes. In fact, we consider a further simplification, which is to replace the finite difference in~\eqref{eq:flowC} by the second derivative in $\omext$, yielding
\begin{equation}    
	\p_s C_{\kappa}(\omext, p) = 
    \mathcal{I}_{\kappa} \p^2_{\omext} C_{\kappa}(\omext, p) \,,
	\label{eq:largep}
\end{equation} 
where
\begin{equation}    
	\mathcal{I}_{\kappa} = 
	\frac{p^2}{2}\int_{\omint,q} q^2 \tilde\p_s C_{\kappa}(\omint, q) =
	\frac{p^2 \kappa D_\kappa}{\nu_\kappa} \int_{\homint,\hq}  
	\frac{\hq^2}{\hat P_\kappa(\homint,\hq)}
	\left( 
		1-
		\frac{2 \hq^4 \hat{k}_\kappa(\homint,\hq)^2}{\hat P_\kappa(\homint,\hq)}
	\right)
	\widehat{\p_s R_\kappa}(\hq)\,,
	\label{eq:largepIk}
\end{equation} 
This assumption is justified by the fact that the frequency integral in~\eqref{eq:flowC} is dominated by  $\omega \ll \omext$. Let us note that here, the $\homint$-dependence of $\hat f_\kappa$ is not neglected.
 For a given $p$, the equation~(\ref{eq:largep}) is a diffusion equation
\begin{equation}
	\p_{\tau} C_{\tau}(\omext) = \mathcal{D}_\tau \p^2_{\omext} C_{\tau}(\omext)    
	\label{eq:largep1}
\end{equation}
with a RG time-dependent diffusion coefficient $\mathcal{D}_\tau \equiv -\mathcal{I}_{\kappa=\Lambda\exp(-\tau)}$, where the RG time variable has been changed to $\tau \equiv -s >0$.
 The numerical scheme to integrate it is presented in Sec.~\ref{sec:num}.

 Prior to this, let us make a comment on the convergence of this equation.
 The numerical results in the small-$p$ regime show that for the KPZ equation the integral $\mathcal{I}_{\kappa}$ is positive for all $\kappa$ and tends to zero as $\kappa\rightarrow0$, $\tau\rightarrow\infty$ (approaching the KPZ fixed point). Thus, the diffusion coefficient $\mathcal{D}_\tau$ is negative. However, this equation is only valid in a specific time domain and should be integrated only in this domain. Indeed, replacing the finite difference by a derivative is not valid for arbiratry $\omext$, such that one has to introduce a cutoff in time  to be consistent.

\section{Numerical scheme for the integration of the FRG flow equations}
\label{sec:num}

In this section, we describe a numerical scheme to integrate together the large-$p$ and  small-$p$ flow equations.
The small-$p$ equation~(\ref{eq:NLOdimless}) is expressed in a dimensionless form.
 Thus, the function $\hat f_\kappa(\homext, \hat p)$ has to be discretised on a dimensionless momentum and frequency grid $(\hp, \homext)$. The corresponding dimensionful momenta $p=\hp\kappa$ decrease with the RG flow as $\kappa$ tends to zero ($\hp$-grid is fixed), therefore only the IR regime is accessible on this grid.
 To describe the UV regime, one should also consider a grid of fixed dimensionful momenta, where
  the RG evolution is governed by the large-$p$ equation.

\begin{figure}
	\centering
    \includegraphics[width=7.2cm]{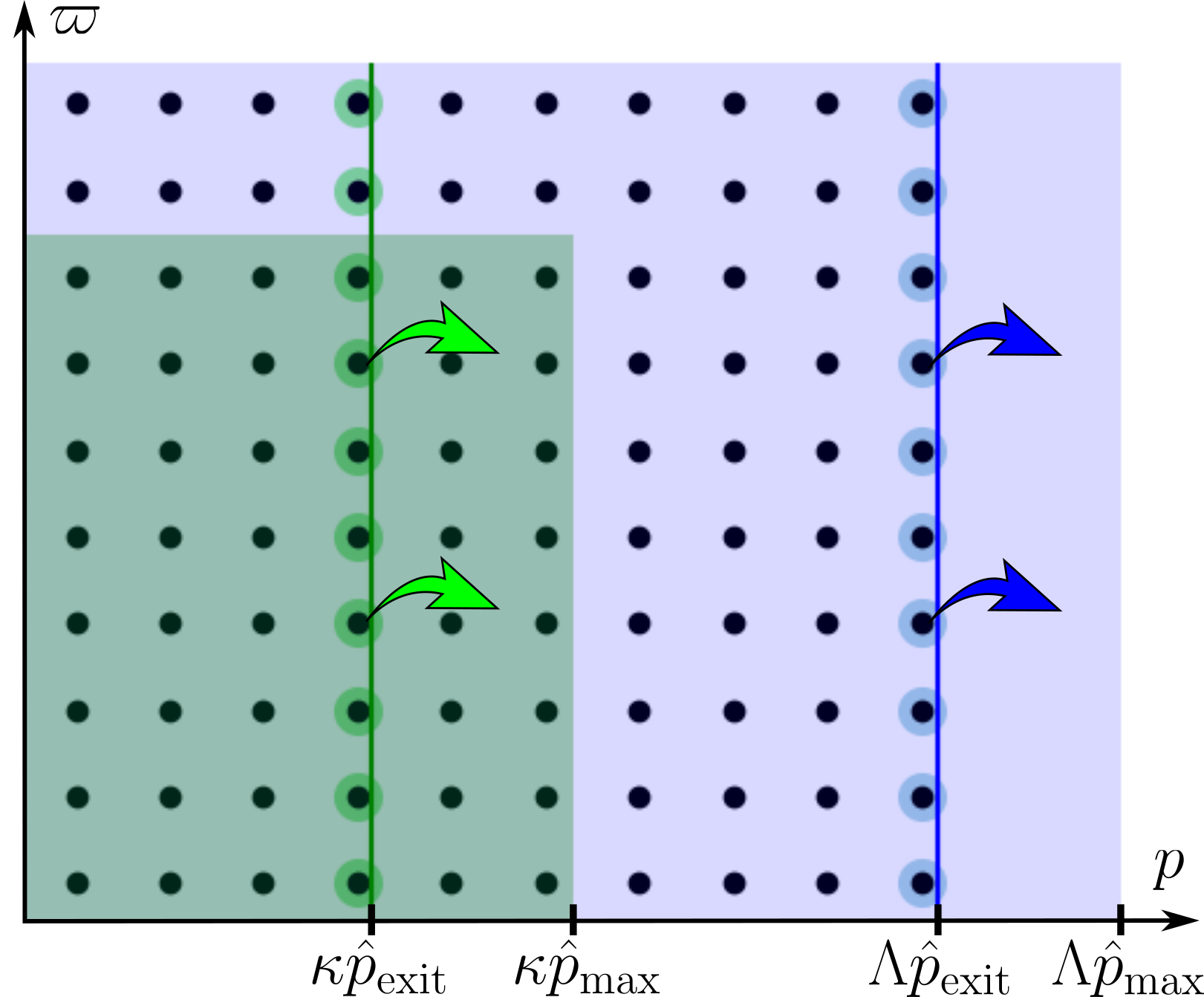}
    \caption{
         Sketch of the two grids. The blue area depicts the dimensionless grid at $\kappa=\Lambda$; the black dots symbolise the nodes of the dimensionful grid, the dots highlighted with blue are about to exit the small-$p$ grid and start the large-$p$ evolution. The green area depicts the dimensionless grid at $\kappa<\Lambda$, and the dots highlighted with green will exit at this value of $\kappa$.
    }
    \label{fig:twogrids_scheme}
\end{figure}

 This can be achieved adapting the scheme proposed in~\cite{Blaizot2006,Benitez2009,Mathey2017},
 as is summarised in Fig.~\ref{fig:twogrids_scheme}. It relies on the use of two grids: a
 dimensionless grid $\{(\hp_i,\homext_j)\}_{i,j=0}^{N_p,N_{\omext}}$ (measured in the flowing units of $\kappa$) and a dimensionful one $\{(\ptwo_i,\omtwo_j)\}_{i,j=0}^{N_{\ptwo},N_{\omtwo}}$  (measured in fixed units, $\Lambda$ for example). The small-$p$ equations are solved on the dimensionless grid, following standard methods described below. Besides, we introduce an \textit{exit criterion}: when $\kappa$ becomes small enough that one of the points of the dimensionful grid $\ptwo_{i_*}$ falls outside of some exit threshold, that is, $\ptwo_{i_*} \geq \kappa\hpexit$, the function $C_\kappa(\forall\omtwo_j,\ptwo_{i_*})$ stops the small-$p$ evolution~(\ref{eq:NLOdimless}) and starts the  large-$p$ evolution according to the equation~(\ref{eq:largep}), which is integrated numerically on the dimensionful grid.
Note that the dimensionless grid shrinks also in the $\omext$-direction as $\omext \equiv (\nu_\kappa \kappa^2)\homext$, but we assume that the small-$p$ equations are suitable for any $\omext$. Thus, at a given $p$ and any $\omext$, the function $C_\kappa(\omext,p)$ evolves according to the small-$p$ equation until $p$ exits. After that, it evolves according to the large-$p$ equation at any $\omext$.

In this work, we use logarithmic dimensionless grids $\{(\hp_i,\homext_j)\}_{i,j=0}^{N_p,N_{\omext}}$ with an additional zero point. The typical parameters are $\hp_{\min}\equiv\hp_1 = 0.01$, $\hp_{\max}\equiv\hp_{N_p} = 100$, $\homext_{\min} = \hp_{\min}^2$, $\homext_{\max} = \hp_{\max}^2$, $N_{\omext} = N_p = 49$. 
In order to perform numerical integration in frequency and momentum, we also introduce dimensionless Gauss–Legendre grids $\{\hq_i\}_{i=0}^{N_q}$ and $\{\homint_j\}_{j=0}^{N_{\omint}}$ with $\hq_{\max} \approx 10$, $N_q=39$ and $\homint_{\max} \approx \homext_{\max}$, $N_{\omint}=499$. The choice of the upper limit in $\hq$ is justified by the fact that all the integrations over momentum include a regulator multiplier, which ensures fast decay of the integrand. The integration over frequency does not share this feature, and the upper limit, as well as $N_q$ and $N_{\omint}$, were chosen empirically to maintain the desired precision (4 digits).
The NLO equation~(\ref{eq:NLOdimless}) is solved using the explicit Euler method with a RG-time step $\Delta s$ varying from $0.001$ for smaller $\hat g_{\Lambda}$ to $0.0001$ for larger $\hat g_{\Lambda}$. 
The diffusion coefficient $\mathcal{D}_\tau$ and the initial condition $C_{\text{IC}}(\omtwo,\ptwo_{i}) \equiv C_{\kappa=\ptwo_{i}/\hpexit}(\omtwo,\ptwo_{i})$ for each $\ptwo_{i}$ are calculated and recorded during the small-$p$ evolution and are inputs to integrate the large-$p$ equation. To construct the correlation function and perform the numerical integration, a power-law continuation of $\hat f(\homext,\hp)$ in $\homext$ is used where necessary.

The dimensionful grid $\{(\ptwo_i,\omtwo_j)\}_{i,j=0}^{N_{\ptwo},N_{\omtwo}}$ is logarithmic as well, with $\ptwo_{\max}\equiv\Lambda\hpexit$, $\omtwo_{\max}\equiv\ptwo_{\max}^2$. The choice of $\hpexit$ is discussed below.
The large-$p$ equation~(\ref{eq:largep}) is integrated using the Crank–Nicolson algorithm. 
 As mentioned earlier, it is a diffusion equation with a negative RG time-dependent diffusion coefficient, but it can  nevertheless be integrated. Due to the fast decay of the diffusion coefficient with RG time (note that {$\mathcal{D}_\tau \sim \kappa^3 \rightarrow 0$}) the solution does not diverge, provided that $\hpexit$ is large enough, as we explain below.

 The meaning of $\hpexit$ is to distinguish between the regions of validity of the small-$p$ and the large-$p$ approximations, but the choice of the specific value remains somewhat arbitrary. The large-$p$ region can be defined according to the shape of the regulator: $10^{-11}$ decay of $\hat r(\hq)$ compared to $\hat r(\hq_{\min})$ is achieved at $\hq\approx4$, so $\hpexit$ can be chosen close to this value. The larger value of $\hpexit$ we choose, the longer RG time the correlation function evolves according to the small-$p$ equations, and the shorter time it spends under the large-$p$ evolution. Thus, the solution should be more stable when using a larger $\hpexit$.
For example, for $\hat g_{\Lambda} = 300$, the Crank–Nicolson algorithm of~(\ref{eq:largep}) leads to instabilities for several values of $\ptwo$ if we choose $\hpexit = 10$ while with $\hpexit = 12.5$ we were able to obtain the solution. If we choose a very large $\hpexit$, say, $\hpexit=\hp_{\max}$, the whole RG flow is governed by the small-$p$ equations, which is not correct since the NLO approximation is derived for small momenta. Surprisingly, even this choice of $\hpexit$ gives reasonable results: the initial condition $C_{\text{IC}}(\omtwo,\ptwo_{i})$ exhibits the expected scaling behaviour, and the large-$p$ evolution does not give a noticeable change in this case.
Our strategy is to choose  a $\hpexit$ such that, on the one hand, the solution of the diffusion equation is stable, and on the other hand, the large-$p$ evolution gives a noticeable change. The convergence of the solution was checked, and it is quite fast -- already at $|s|\sim 5$.

\section{Results}
\label{sec:res}

We integrated the equations~(\ref{eq:NLOdimless}),(\ref{eq:largep}) as described in the previous section, and recorded the correlation function $C(\omtwo,\ptwo)$ defined by ~(\ref{eq:Corr}) at the RG time of exiting of each $\ptwo$ before the large-$p$ evolution (which serves as initial condition $C_{\text{IC}}$ for~(\ref{eq:largep})) and after the large-$p$ equation integration. We present below the results, starting with the identification of the different scaling regimes through the half-frequency, then displaying the correlation functions, and finally discussing the extent of the IB regime.

\subsection{Scaling regimes}

\begin{figure}
	\centering
    \includegraphics[width=7.2cm]{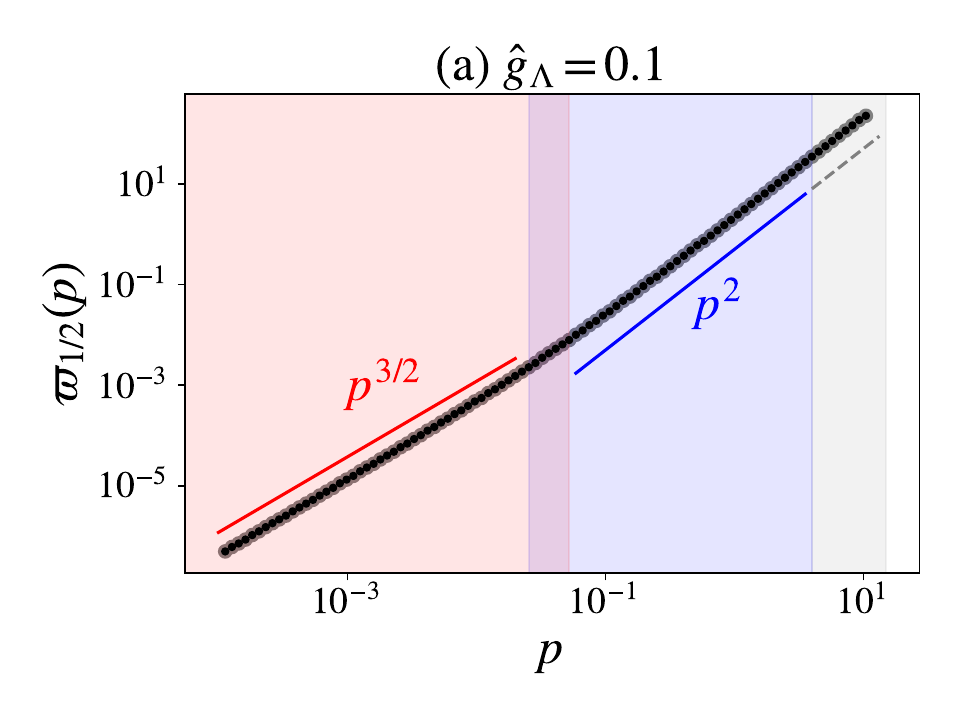}
    \includegraphics[width=7.2cm]{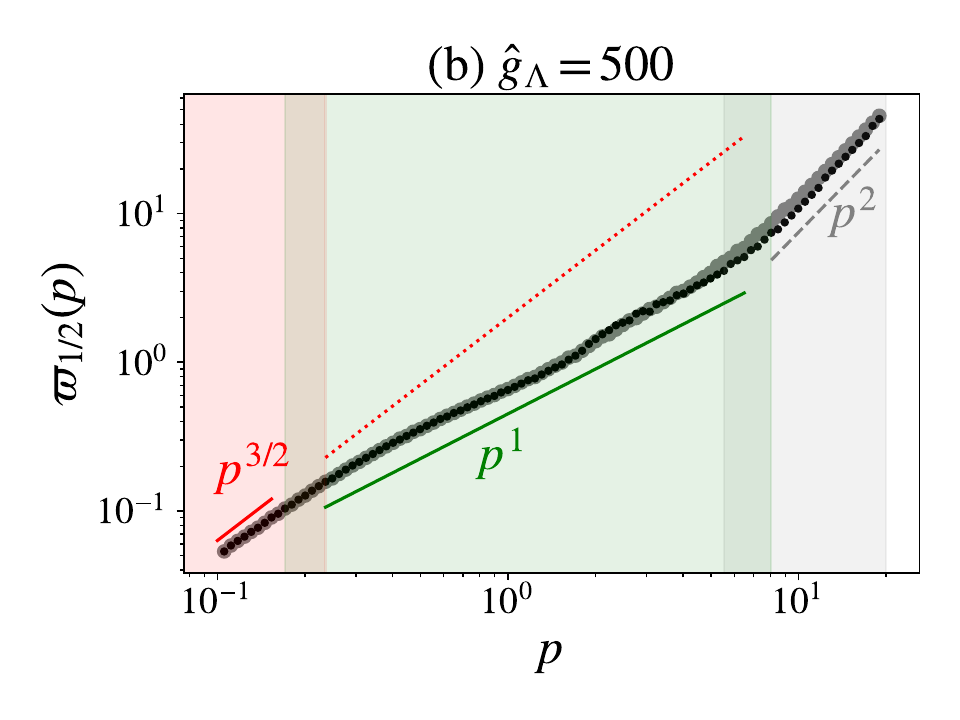}
    \caption{
         Half-frequency $\omext_{1/2}(p)$ as a function of momentum $p$ calculated before (gray dots) and after (black dots) the large-$p$ evolution given by~(\ref{eq:largep}),  at (a)~$\hat g_{\Lambda} = 0.1$ and (b)~$\hat g_{\Lambda} = 500$. The different shades show the IB (green), KPZ (red) and EW (blue) universal scaling regimes. The plain lines are guidelines showing the expected power-laws. The red dotted line with $3/2$ slope in the intermediate region of panel (b) is provided for comparison.
    }
    \label{fig:scaling}
\end{figure}

The various scaling regimes can be identified by defining the half-frequency $\omext_{1/2}(p)$, which is given by the relation
\begin{equation}
	C(\omext_{1/2}(p),p) = C(0,p)/2\,.
\label{eq:omegahalf}
\end{equation}
In a scaling regime, it is expected to behave as $\omext_{1/2}(p)\sim p^z$ where $z$ is the dynamical critical exponent.
The result is displayed in Fig.~\ref{fig:scaling} for two values of the bare coupling $\hat g_\Lambda$.
 For a small value $\hat g_\Lambda$, corresponding to a small non-linearity $\lambda$ or equivalently a large viscosity $\nu$, one expects two universal scaling regimes: the KPZ one with  $z=3/2$ in the IR and the EW one with  $z=2$ in the UV, which is what is observed  in Fig.~\ref{fig:scaling}(a), in the red-shaded, and {blue}-shaded regions respectively. For the largest momenta (gray shade), the result is non-universal, it merely reflects the initial condition. In this example, the initial condition has the bare scaling $p^2$, which is similar to the EW one, thus it is undistinguishable from the former in this figure. However, we also implemented other initial conditions with different behaviours, and in these cases, the initial behaviours are imprinted in the gray region instead of the $p^2$ scaling, whereas in the blue intermediate UV regime, the curve is unchanged and still displays an EW scaling. This regime is thus universal, controlled by the  EW fixed point.
 In the IR, the KPZ regime is reached for any initial condition.  For small values of $\hat g_\Lambda$, the large-$p$ evolution gives a negligible change with respect to  the initial condition $C_{\text{IC}}$. Indeed, for small values of the non-linear coupling,  the r.h.s. of the large-$p$ equation~(\ref{eq:largep}) is very small. Let us emphasise that this situation corresponds to the typical one studied by mathematicians, who focus on the trajectories from the EW to the KPZ fixed point~\cite{Corwin12}.

An example of  $\omext_{1/2}(p)$  for a large value of $\hat g_\Lambda$, corresponding to large non-linearity, or equivalently  small viscosity, is shown in Fig.~\ref{fig:scaling}(b).
In contrast, this situation has not been investigated yet by mathematicians, as it corresponds to trajectories to the KPZ fixed point from higher, rather than smaller, values of the coupling. While the KPZ regime with $z=3/2$ is still visible in the IR, it is now superceded by the IB regime, with $z=1$, instead of the EW one in the intermediate region. Even if the $z=1$ regime is numerically more noisy, it is unambiguously identified in Fig.~\ref{fig:scaling}(b), and it clearly deviates from the  IR scaling $z=3/2$  also shown for comparison.
 Note that the largest momenta (gray-shaded region) still display a $p^2$ scaling, which  corresponds to the non-universal region, determined by the initial condition,  discussed  in the previous case. We also implemented in this case different initial conditions, with the same observation: the gray region is changed following the initial condition, whereas the intermediate green region remains unaltered, this regime is universal.
Thus, it shows that in the limit of vanishing viscosity, another UV fixed point, the IB one, emerges and controls the universal large momentum behaviour of the correlation function. 
Let us comment on the KPZ regime in this figure. For any  values of $\hat g_\Lambda$, the flow reaches the KPZ fixed point as $\kappa\rightarrow0$  ($s\rightarrow-\infty$), which explains the appearance of the KPZ scaling in the IR. However, for the large $g_\Lambda$, we stopped the RG flow at a relatively small RG time $|s|\sim6$ (when the last point $\ptwo=\ptwo_{\min}$ exits) while the KPZ fixed point  is rather expected at larger times $|s|\sim15$~\cite{Fontaine2023InvBurgers}, therefore the KPZ regime is not yet well developed even at $\ptwo_{\min}$ in Fig.~\ref{fig:scaling}(b).

\subsection{Correlation function}

\begin{figure}
	\centering
    \includegraphics[width=7.2cm]{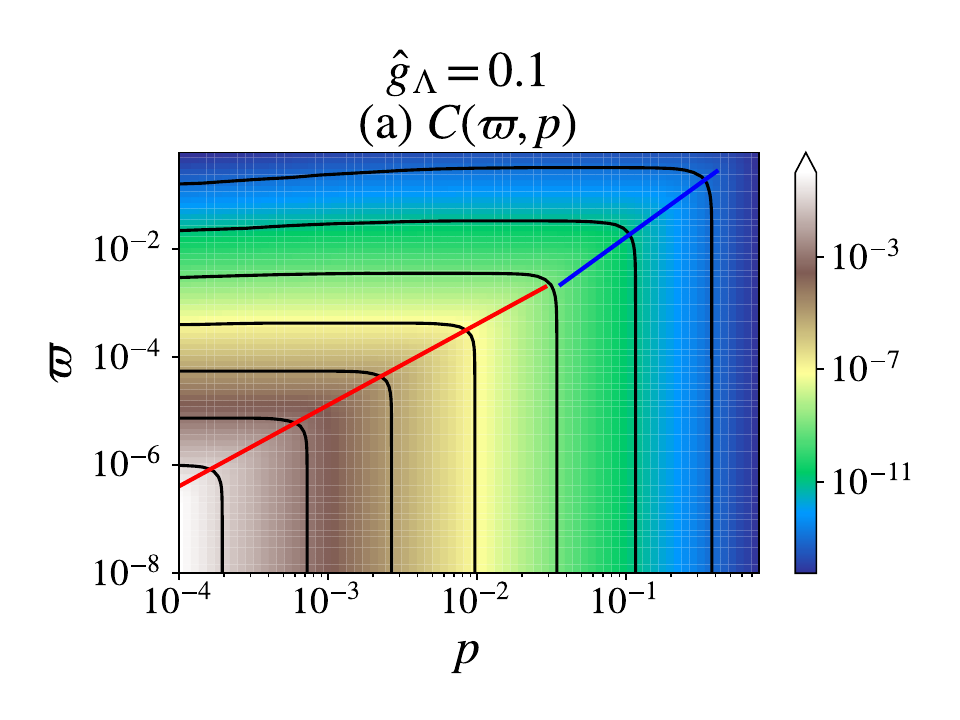}
    \includegraphics[width=7.2cm]{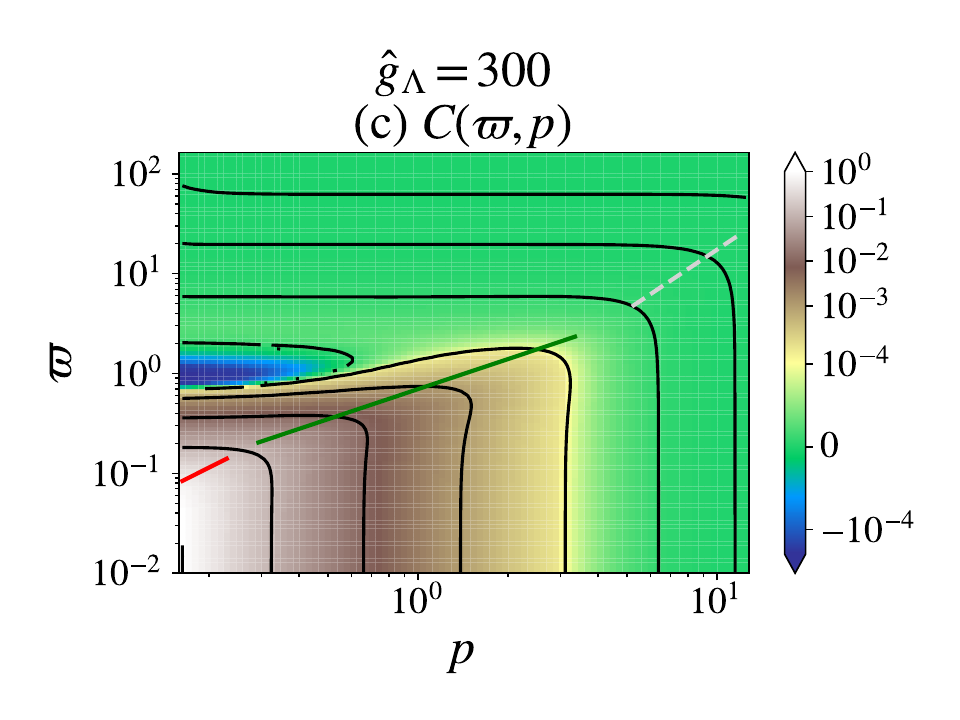}
    \includegraphics[width=7.2cm]{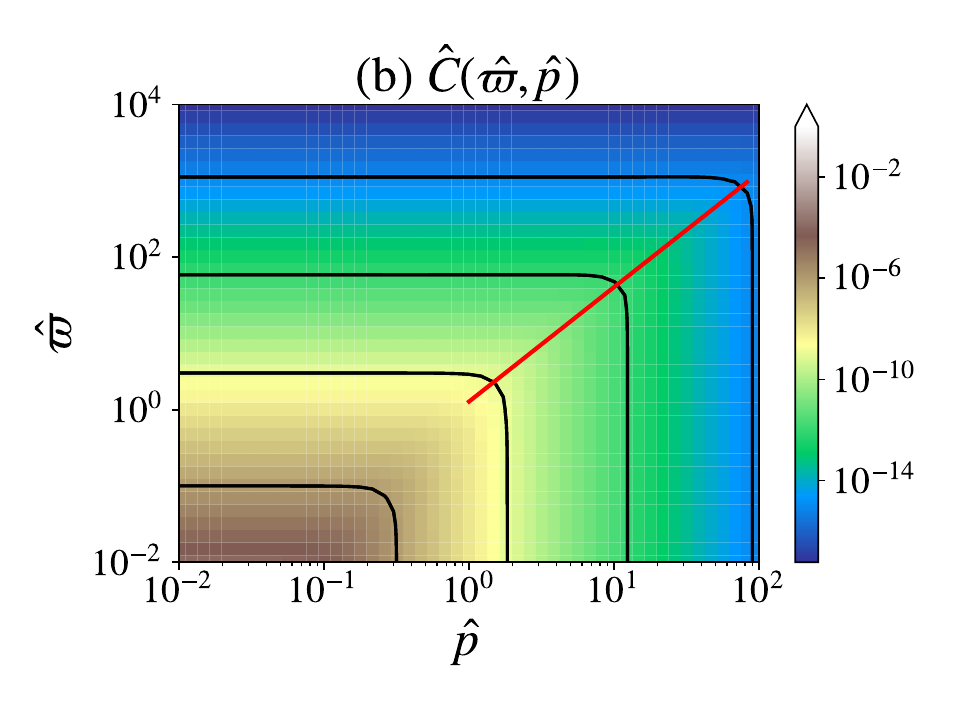}
    \includegraphics[width=7.2cm]{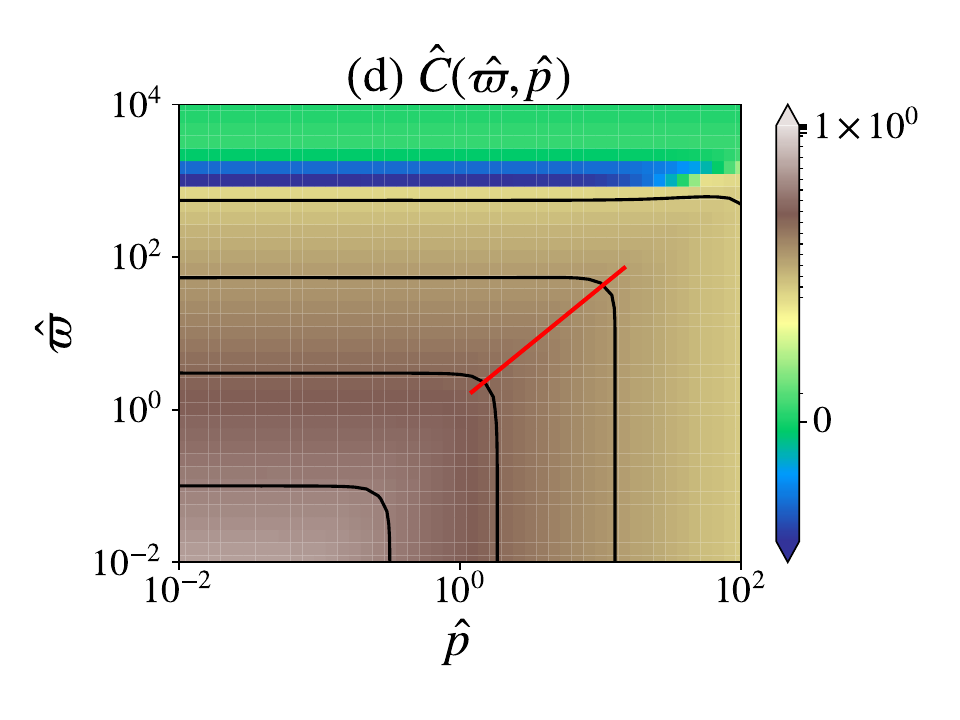}
    \caption{
         (a)  dimensionful  and (b) dimensionless correlation functions at small $\kappa$ with $\hat g_{\Lambda} = 0.1$; (c) dimensionful  and (d) dimensionless correlation functions for $\hat g_{\Lambda} = 300$. The lines show the IB (green), KPZ (red) and EW (blue) scaling regimes, the dashed gray line shows the non-universal EW-like scaling.
    }
    \label{fig:heatmap}
\end{figure}

The two situations of small and large initial $g_\Lambda$ yield very different physical properties.
As an example, the recorded correlation function $C_{\text{IC}}$ are shown in Fig.~\ref{fig:heatmap} for a small (a) and a large (c) value of $g_\Lambda$. Both IR and UV regimes can be observed, but  they have very different shapes, emphasised by the contour lines. In particular, the IB regime features non-trivial structures, with a negative dip, which is absent in the EW regime. We also show in Fig.~\ref{fig:heatmap}(b) and (d), the corresponding dimensionless correlation function $\hat C(\homext, \hp)$ constructed with the NLO functions $\hat f_\kappa(\hat\varpi,\hat p)$
recorded at one particular $\kappa$ in the IR (corresponding to the exit of the last point $\ptwo=\ptwo_{\min}$). On these maps, one can  mainly observe the KPZ scaling: at this $\kappa$ the flow is already close enough to the attractive KPZ fixed point (still having a negative dip in the $\hat g_{\Lambda}=300$ case -- a reminiscence of the IB region). These plots show that the dimensionful grid is necessary to well capture the UV behaviour of the functions.

Let us note that in the IB region and the transition region the correlation function $C(\omext,p)$ has a hump at $\omext>0$ (that is, $\max_{\omext}C(\omext,p) \neq C(0,p)$). This suggests a possible re-definition of the half-frequency~(\ref{eq:omegahalf}) as $C(\omext_{1/2}(p),p) = \max_{\omext}C(\omext,p)/2$\,.
However, we checked that with this definition,  the curves in Fig.~\ref{fig:region} are just uniformly shifted, enlarging the IB region roughly by a factor of two.

\subsection{Width of the IB regime}

\begin{figure}
	\centering
    \includegraphics[width=7.2cm]{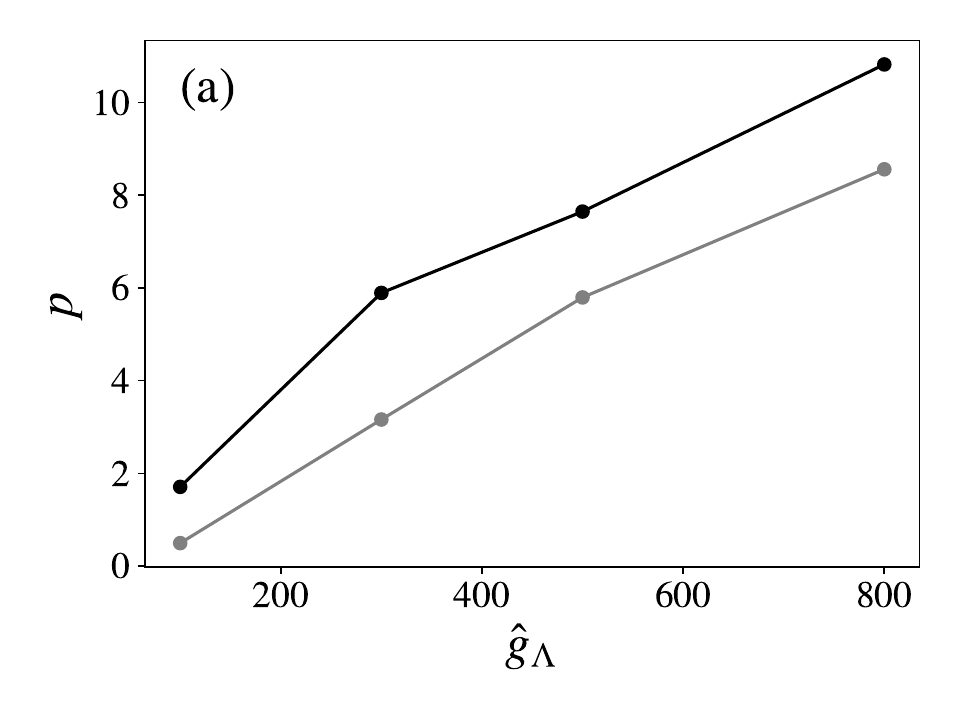}
    \includegraphics[width=7.2cm]{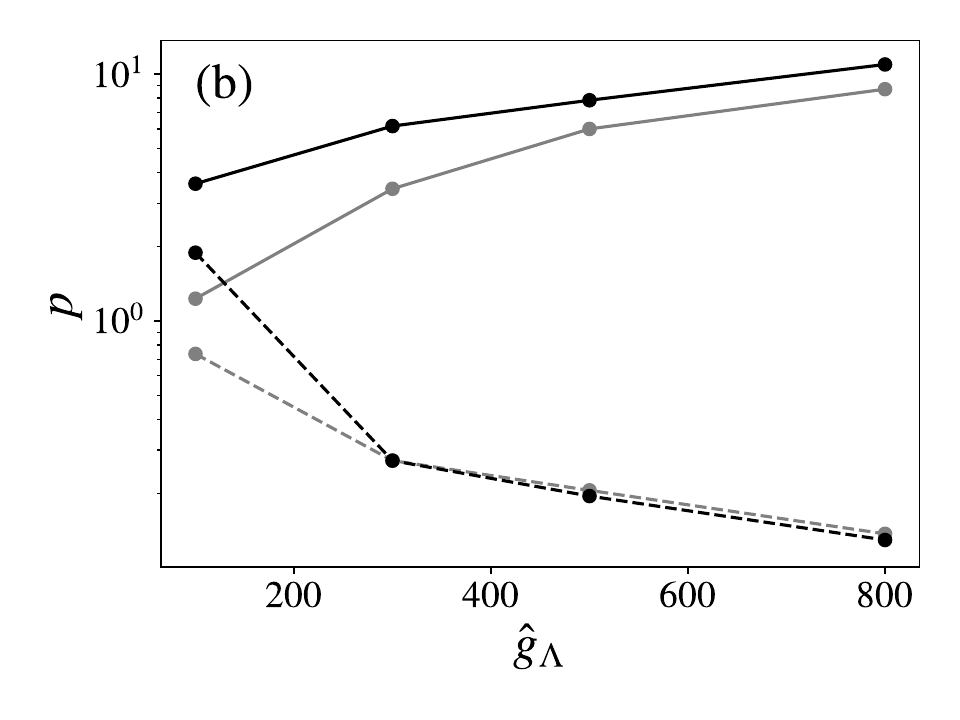}
    \caption{Width of the IB region (a) and its upper and lower bounds (b) as functions of the microscopic coupling $\hat g_{\Lambda}$ calculated before (gray dots) and after the large-$p$ evolution given by Eq.~(\ref{eq:largep}) (black dots).}
    \label{fig:region}
\end{figure}

This coupled integration of the large-$p$ and small-$p$ equations allows us to make a quantitative study of the extent of the IB regime as a function of the viscosity, or equivalently of $g_\Lambda$.
We performed calculations of the correlation function for several values of $\hat g_{\Lambda}$ from $100$ to $800$, and determined for each the half-frequency $\omext_{1/2}(p)$. At $\hat g_{\Lambda}=100$ the IB regime is very narrow (less than one decade in momentum, both for the initial condition $C_{\text{IC}}$ and the large-$p$  solution), which provides an estimation of a threshold value above which the coupling can be considered as ``large'' and the IB scaling manifests itself. The IB region extends as $\hat g_{\Lambda}$ grows. To quantify this growth, we define a region with a given exponent $z$  as the widest region of $p$ where the slope of a linear fit of $\ln\omext_{1/2}$ {\it vs} $\ln p$ lies within $z \pm 4\%$. From these results, we identify the lower bound and the upper bound of the IB region (delineating it from the KPZ one and the non-universal  one respectively), and deduce the width of this region. The results are displayed in Fig.~\ref{fig:region}, both  for the initial condition and after the large-$p$ evolution. The lower limit is almost the same for the initial condition $C_{\text{IC}}$ and the large-$p$  solution. Moreover, it does not depend on the choice of the microscopic initial condition $\hat f_{\Lambda}(\homext,\hp)$ for the NLO equation, as expected: by the time the flow reaches the vicinity of the KPZ fixed point, it has erased its initial condition. The upper bound is shifted upwards after the large-$p$ evolution, enlarging the IB region.
Thus the large-$p$ equation provides  some improvement over the small-$p$ equation to describe the IB region, although this regime is surprisingly already observable at the NLO level.

\section{Conclusion}
\label{sec:conclusion}

Within the FRG formalism, two complementary approximation schemes were proposed to study the KPZ-Burgers equation. One was used  to describe the small momentum IR regime, the other was devised to access the large momentum UV sector, in both cases compared to the RG scale $\kappa$. The two approaches were so far studied independently.
In this work,  we have proposed a numerical scheme to integrate simultaneously the two sets of flow equations, referred to as small-$p$ and large-$p$ ones. This scheme consists in solving the small-$p$ equations on a dimensionless frequency-momentum grid, which is used to progressively  initialise the large-$p$ evolution on a separate dimensionful frequency-momentum grid. The large-$p$ evolution is  then integrated, using as inputs quantities computed on the small-$p$ grid.

The advantage of this scheme is to allow one to obtain a full unified picture of the properties of the system at all scales, encompassing both its IR and its UV behaviour.
In particular, we have computed the correlation function, and the associated half-frequency $\omext_{1/2}(p)$ at which it has decayed to half of its value at $\varpi=0$ for each $p$.
We have investigated the effect of the initial parameter $g_\Lambda = \lambda^2 D/\nu^3$, and evidenced the crossover in the UV from the EW scaling regime at small $g_\Lambda$ characterised by $\omext_{1/2}(p)\sim p^2$ to the IB scaling regime at large $g_\Lambda$ ({\it i.e.} in the inviscid limit) characterised by $\omext_{1/2}(p)\sim p$, while the IR is always controlled by the KPZ scaling. We have determined the width in momenta of the IB region as a function of  $g_\Lambda$, and showed that it increases approximately linearly.
 In perspective, this scheme can be easily adapted and used to study different systems. It would be very useful for the Navier-Stokes equation, to study the crossover between different scaling regimes which can be realised when changing the viscosity, from a near-equilibrium situation to a fully turbulent one.

\section*{Acknowledgements}
This work received the support of the French-Uruguayan Institute
of Physics (IFU$\Phi$). LG acknowledges support by the MSCA Cofund QuanG (Grant Number : 101081458) funded by the European Union. Views and opinions expressed are however those of the authors only and do not necessarily reflect those of the European Union or Université Grenoble Alpes. Neither the European Union nor the granting authority can be held responsible for them.

\nolinenumbers

\end{document}